\documentclass[twocolumn]{aastex62}
\usepackage{graphicx}
\usepackage{grffile}
\usepackage{color}
\usepackage{amsmath}
\usepackage{booktabs}
\usepackage{pstricks}
\usepackage{lineno}

\RequirePackage{flushend}

\newcommand{\chb}{$b\bar{b}$}
\newcommand{\chW}{$W^{+}W^{-}$}
\newcommand{\chtau}{$\tau^{+}\tau^{-}$}

\begin{document}

\title{Velocity independent constraints on spin-dependent DM-nucleon interactions from IceCube and PICO.}

\author{M. G. Aartsen}
\affiliation{Dept. of Physics and Astronomy, University of Canterbury, Private Bag 4800, Christchurch, New Zealand}
\author{M. Ackermann}
\affiliation{DESY, D-15738 Zeuthen, Germany}
\author{J. Adams}
\affiliation{Dept. of Physics and Astronomy, University of Canterbury, Private Bag 4800, Christchurch, New Zealand}
\author{J. A. Aguilar}
\affiliation{Universit{\'e} Libre de Bruxelles, Science Faculty CP230, B-1050 Brussels, Belgium}
\author{M. Ahlers}
\affiliation{Niels Bohr Institute, University of Copenhagen, DK-2100 Copenhagen, Denmark}
\author{M. Ahrens}
\affiliation{Oskar Klein Centre and Dept. of Physics, Stockholm University, SE-10691 Stockholm, Sweden}
\author{C. Alispach}
\affiliation{D{\'e}partement de physique nucl{\'e}aire et corpusculaire, Universit{\'e} de Gen{\`e}ve, CH-1211 Gen{\`e}ve, Switzerland}
\author{K. Andeen}
\affiliation{Department of Physics, Marquette University, Milwaukee, WI, 53201, USA}
\author{T. Anderson}
\affiliation{Dept. of Physics, Pennsylvania State University, University Park, PA 16802, USA}
\author{I. Ansseau}
\affiliation{Universit{\'e} Libre de Bruxelles, Science Faculty CP230, B-1050 Brussels, Belgium}
% ...
\date{\today}
\collaboration{IceCube Collaboration}
\thanks{Email: analysis@icecube.wisc.edu}

\author{C.~Amole}
\affiliation{Department of Physics, Queen's University, Kingston, K7L 3N6, Canada}

\author{M.~Ardid}
\affiliation{Departament de F\'isica Aplicada, IGIC - Universitat Polit\`ecnica de Val\`encia, Gandia 46730 Spain}

\author{I.~J.~Arnquist}
\affiliation{Pacific Northwest National Laboratory, Richland, Washington 99354, USA}

\author{D.~M.~Asner}
 \altaffiliation[now at ]{Brookhaven National Laboratory}
\affiliation{Pacific Northwest National Laboratory, Richland, Washington 99354, USA}

\author{D.~Baxter}
\affiliation{
Department of Physics and Astronomy, Northwestern University, Evanston, Illinois 60208, USA}
\affiliation{Enrico Fermi Institute, KICP and Department of Physics, 
University of Chicago, Chicago, Illinois 60637, USA}

\author{E.~Behnke}
\affiliation{Department of Physics, Indiana University South Bend, South Bend, Indiana 46634, USA}




\author{M.~Bressler}
\affiliation{Department of Physics, Drexel University, Philadelphia, Pennsylvania 19104, USA}

\author{B.~Broerman}
\affiliation{Department of Physics, Queen's University, Kingston, K7L 3N6, Canada}


\author{G.~Cao}
\affiliation{Department of Physics, Queen's University, Kingston, K7L 3N6, Canada}

\author{C.~J.~Chen}
\affiliation{
Department of Physics and Astronomy, Northwestern University, Evanston, Illinois 60208, USA}

\author{U.~Chowdhury}
 \altaffiliation[now at ]{Canadian Nuclear Laboratories}
\affiliation{Department of Physics, Queen's University, Kingston, K7L 3N6, Canada}

\author{K.~Clark}
\affiliation{Department of Physics, Queen's University, Kingston, K7L 3N6, Canada}

\author{J.~I.~Collar}
\affiliation{Enrico Fermi Institute, KICP and Department of Physics,
University of Chicago, Chicago, Illinois 60637, USA}

\author{P.~S.~Cooper}
\affiliation{Fermi National Accelerator Laboratory, Batavia, Illinois 60510, USA}

\author{M.~Crisler}
\affiliation{Fermi National Accelerator Laboratory, Batavia, Illinois 60510, USA}
\affiliation{Pacific Northwest National Laboratory, Richland, Washington 99354, USA}

\author{G.~Crowder}
\affiliation{Department of Physics, Queen's University, Kingston, K7L 3N6, Canada}

\author{N.A.~Cruz-Venegas}
\affiliation{Instituto de F\'isica, Universidad Nacional Aut\'onoma de M\'exico, M\'exico D.\:F. 01000, M\'exico}

\author{C.~E.~Dahl}
\affiliation{
Department of Physics and Astronomy, Northwestern University, Evanston, Illinois 60208, USA}
\affiliation{Fermi National Accelerator Laboratory, Batavia, Illinois 60510, USA}

\author{M.~Das}
\affiliation{Astroparticle Physics and Cosmology Division, Saha Institute
of Nuclear Physics, Kolkata, India}

\author{S.~Fallows}
\affiliation{Department of Physics, University of Alberta, Edmonton, T6G 2E1, Canada}

\author{J.~Farine}
\affiliation{Department of Physics, Laurentian University, Sudbury, P3E 2C6, Canada}

\author{I.~Felis}
\affiliation{Departament de F\'isica Aplicada, IGIC - Universitat Polit\`ecnica de Val\`encia, Gandia 46730 Spain}

\author{R.~Filgas}
\affiliation{Institute of Experimental and Applied Physics, Czech Technical University in Prague, Prague, Cz-12800, Czech Republic}

\author{F.~Girard}
\affiliation{Department of Physics, Laurentian University, Sudbury, P3E 2C6, Canada}
\affiliation{D\'epartement de Physique, Universit\'e de Montr\'eal, Montr\'eal, H3C 3J7, Canada}

\author{G.~Giroux}
\affiliation{Department of Physics, Queen's University, Kingston, K7L 3N6, Canada}

\author{J.~Hall}
\affiliation{SNOLAB, Lively, Ontario, P3Y 1N2, Canada}

\author{C.~Hardy}
\affiliation{Department of Physics, Queen's University, Kingston, K7L 3N6, Canada}

\author{O.~Harris}
\affiliation{Northeastern Illinois University, Chicago, Illinois 60625, USA}

\author{E.~W.~Hoppe}
\affiliation{Pacific Northwest National Laboratory, Richland, Washington 99354, USA}

\author{M.~Jin}
\affiliation{
Department of Physics and Astronomy, Northwestern University, Evanston, Illinois 60208, USA}

\author{L.~Klopfenstein}
\affiliation{Department of Physics, Indiana University South Bend, South Bend, Indiana 46634, USA}

\author{C.~B.~Krauss}
\affiliation{Department of Physics, University of Alberta, Edmonton, T6G 2E1, Canada}

\author{M.~Laurin}
\affiliation{D\'epartement de Physique, Universit\'e de Montr\'eal, Montr\'eal, H3C 3J7, Canada}

\author{I.~Lawson}
\affiliation{Department of Physics, Laurentian University, Sudbury, P3E 2C6, Canada}
\affiliation{SNOLAB, Lively, Ontario, P3Y 1N2, Canada}

\author{A.~Leblanc}
\affiliation{Department of Physics, Laurentian University, Sudbury, P3E 2C6, Canada}

\author{I.~Levine}
\affiliation{Department of Physics, Indiana University South Bend, South Bend, Indiana 46634, USA}

\author{W.~H.~Lippincott}
\affiliation{Fermi National Accelerator Laboratory, Batavia, Illinois 60510, USA}

\author{F.~Mamedov}
\affiliation{Institute of Experimental and Applied Physics, Czech Technical University in Prague, Prague, Cz-12800, Czech Republic}

\author{D.~Maurya}
\affiliation{Bio-Inspired Materials and Devices Laboratory (BMDL), Center for Energy Harvesting Material and Systems (CEHMS), Virginia Tech, Blacksburg, Virginia 24061, USA}

\author{P.~Mitra}
\affiliation{Department of Physics, University of Alberta, Edmonton, T6G 2E1, Canada}

\author{C.~Moore}
\affiliation{Department of Physics, Queen's University, Kingston, K7L 3N6, Canada}

\author{T.~Nania}
\affiliation{Department of Physics, Indiana University South Bend, South Bend, Indiana 46634, USA}

\author{R.~Neilson}
\affiliation{Department of Physics, Drexel University, Philadelphia, Pennsylvania 19104, USA}

\author{A.~J.~Noble}
\affiliation{Department of Physics, Queen's University, Kingston, K7L 3N6, Canada}

\author{P.~Oedekerk}
\affiliation{Department of Physics, Indiana University South Bend, South Bend, Indiana 46634, USA}


\author{A.~Ortega}
\affiliation{Enrico Fermi Institute, KICP and Department of Physics,
University of Chicago, Chicago, Illinois 60637, USA}

\author{M.-C.~Piro}
\affiliation{Department of Physics, University of Alberta, Edmonton, T6G 2E1, Canada}

\author{A.~Plante}
\affiliation{D\'epartement de Physique, Universit\'e de Montr\'eal, Montr\'eal, H3C 3J7, Canada}

\author{R.~Podviyanuk}
\affiliation{Department of Physics, Laurentian University, Sudbury, P3E 2C6, Canada}

\author{S.~Priya}
\affiliation{Bio-Inspired Materials and Devices Laboratory (BMDL), Center for Energy Harvesting Material and Systems (CEHMS), Virginia Tech, Blacksburg, Virginia 24061, USA}

\author{A.~E.~Robinson}
\affiliation{D\'epartement de Physique, Universit\'e de Montr\'eal, Montr\'eal, H3C 3J7, Canada}



\author{S.~Sahoo}
\affiliation{Astroparticle Physics and Cosmology Division, Saha Institute
of Nuclear Physics, Kolkata, India}

\author{O.~Scallon}
\affiliation{Department of Physics, Laurentian University, Sudbury, P3E 2C6, Canada}

\author{S.~Seth}
\affiliation{Astroparticle Physics and Cosmology Division, Saha Institute
of Nuclear Physics, Kolkata, India}

\author{A.~Sonnenschein}
\affiliation{Fermi National Accelerator Laboratory, Batavia, Illinois 60510, USA}

\author{N.~Starinski}
\affiliation{D\'epartement de Physique, Universit\'e de Montr\'eal, Montr\'eal, H3C 3J7, Canada}

\author{I.~\v{S}tekl}
\affiliation{Institute of Experimental and Applied Physics, Czech Technical University in Prague, Prague, Cz-12800, Czech Republic}

\author{T.~Sullivan}
\affiliation{Department of Physics, Queen's University, Kingston, K7L 3N6, Canada}

\author{F.~Tardif}
\affiliation{D\'epartement de Physique, Universit\'e de Montr\'eal, Montr\'eal, H3C 3J7, Canada}

\author{E.~V\'azquez-J\'auregui}
\affiliation{Instituto de F\'isica, Universidad Nacional Aut\'onoma de M\'exico, M\'exico D.\:F. 01000, M\'exico}
\affiliation{Department of Physics, Laurentian University, Sudbury, P3E 2C6, Canada}

\author{N.~Walkowski}
\affiliation{Department of Physics, Indiana University South Bend, South Bend, Indiana 46634, USA}


\author{U.~Wichoski}
\affiliation{Department of Physics, Laurentian University, Sudbury, P3E 2C6, Canada}

\author{Y.~Yan}
\affiliation{Bio-Inspired Materials and Devices Laboratory (BMDL), Center for Energy Harvesting Material and Systems (CEHMS), Virginia Tech, Blacksburg, Virginia 24061, USA}

\author{V.~Zacek}
\affiliation{D\'epartement de Physique, Universit\'e de Montr\'eal, Montr\'eal, H3C 3J7, Canada}

\author{J.~Zhang}
\altaffiliation[now at ]{Argonne National Laboratory}
\affiliation{
Department of Physics and Astronomy, Northwestern University, Evanston, Illinois 60208, USA}

\collaboration{PICO Collaboration}
\thanks{Email: analysis@picoexperiment.com}
\noaffiliation

\begin{abstract}

Adopting the Standard Halo Model (SHM) of an isotropic Maxwellian velocity distribution for dark matter (DM) particles in the Galaxy, the most stringent current constraints on their spin-dependent scattering cross-section with nucleons come from the IceCube neutrino observatory and the PICO-60 C$_3$F$_8$ superheated bubble chamber experiments. The former is sensitive to high energy neutrinos from the self-annihilation of DM particles captured in the Sun, while the latter looks for nuclear recoil events from DM scattering off nucleons. Although slower DM particles are more likely to be captured by the Sun, the faster ones are more likely to be detected by PICO.  Recent N-body simulations suggest significant deviations from the SHM for the smooth halo component of the DM, while observations hint at a dominant fraction of the local DM being in substructures. We use the method of \cite{Ferrer:2015bta} to exploit the complementarity between the two approaches and derive conservative constraints on DM-nucleon scattering. Our results constrain $\sigma_{\mathrm{SD}} \lesssim 3 \times 10^{-39} \mathrm{cm}^2$ ($6 \times 10^{-38} \mathrm{cm}^2$) at $\gtrsim 90\%$  C.L. for a DM particle of mass 1~TeV annihilating into $\tau^+ \tau^-$ ($b\bar{b}$) with a local density of $\rho_{\mathrm{DM}} = 0.3~\mathrm{ GeV/cm}^3$. The constraints scale inversely with $\rho_{\mathrm{DM}}$ and are independent of the DM velocity distribution.

\end{abstract}

\section{Introduction}

Based on inferences from observations of gravitational effects, it has long been believed that a significant fraction of the Universe is made up of dark matter (DM) (see \cite{vandenBergh:1999sa}). However, very little is known about its properties and interactions. A weakly interacting massive particle (WIMP), whose relic abundance from a state of thermal equilibrium can make up DM has been the subject of considerable theoretical attention and experimental focus (see~\cite{Bertone:2004pz} for a comprehensive review).

Various complementary approaches have been pursued to detect the WIMPs that may constitute the DM halo of our Galaxy. Terrestrial direct detection (DD) experiments search for nuclear recoil events from the elastic scattering of WIMPs with the target nuclei of their detectors. Neutrino and gamma ray telescopes search for directional excesses over astrophysical backgrounds that may indicate the pair-annihilation of WIMPs, while collider searches look for the signatures of WIMPs being created in high-energy interactions of Standard Model particles. 

Although the different search strategies have attained the sensitivity to probe the physically-motivated WIMP parameter space over the past few decades, they have failed to detect any signal. In the absence of a convincing detection, constraints have been derived on the interaction cross-sections of these hypothetical particles with Standard Model particles. Such an inference requires knowledge both of the density of DM $\rho_{\mathrm{DM}}$ and of its velocity distribution function (VDF) $f(\vec{v})$. 

In the Standard Halo Model (SHM) \citep{Drukier:1986tm}, the DM of the halo is a collisionless gas in hydrostatic equilibrium with the stars, retaining the velocity distribution obtained during the formation of our Galaxy. An isotropic Maxwell-Boltzman velocity distribution in the Galactic rest frame is usually adopted.  

Meanwhile, N-body simulations have hinted that a Maxwell-Boltzmann distribution does not accurately represent even the smooth component of the halo \citep{Kuhlen:2009vh, Lisanti:2010qx, Mao:2012hf}. Recent observations point to the possibility that a dominant fraction of the DM in the Solar neighbourhood~\citep{Necib:2018iwb} may not yet have achieved dynamical equilibrium, perhaps due to the infalling tidal debris of a disrupted massive satellite galaxy of the Milky Way. New data also suggest that a substantial fraction of our stellar halo may lie in a strongly radially anisotropic population, the `Gaia sausage'~\citep{Evans:2018bqy}.


If so, constraints on WIMP-nucleon interactions derived assuming the SHM (from both direct and indirect searches) may be weakened. Direct detection experiments are preferentially sensitive to nuclear recoils from high velocity DM particles, while capture in the Sun is more likely for the slower fraction of the DM population. In this work we use the method of ~\cite{Ferrer:2015bta}, which is independent of the velocity distribution of the halo model to exploit this complementarity and derive \emph{conservative}, upper limits on the spin-dependent DM-nucleon scattering cross-section by combining the results from \cite{Aartsen:2016zhm} and \cite{Amole:2017dex}. Here the DM velocity distribution is taken to be a completely general superposition of individual 'streams' (delta functions in velocity), similarly to the halo-independent analysis of direct detection experiments ~\citep{Frandsen:2011gi}. Although constraints from individual searches will now be dependent on the stream velocity, by exploiting the complementarity of the IceCube and PICO searches, constraints independent of the stream velocity can be obtained. This method also improves on previous assessments of halo model uncertainties on indirect DM detection ~\citep{Choi:2013eda}, by allowing the velocity distribution to be anisotropic. The resulting constraints are a factor of 2 to 4 worse than the PICO SHM constraints at low DM masses and up to an order of magnitude worse at high DM masses, depending upon the annihilation channel, but are independent of the halo model.

\section{Detectors and data samples}
\label{sec:data}
\subsection{IceCube 3 year Solar WIMP search}

IceCube is a cubic-kilometer neutrino detector installed in the ice at the geographic South Pole between depths of 1450 and 2450 m. It relies on photomultiplier tubes housed in pressure vessels known as digital optical modules (DOM) for the optical detection of Cherenkov photons emitted by charged particles traversing the ice. The principal IceCube array is sensitive to neutrinos down to $\sim$100~GeV in energy \citep{Achterberg:2006md, Abbasi:2008aa, Aartsen:2016nxy}. The central region of the detector is an infill array known as DeepCore optimized in geometry and DOM density for the detection of neutrinos at lower energies, down to $\sim$10~GeV~\citep{Collaboration:2011ym}. 

Over a detector uptime of 532 days corresponding to the austral winters between May 2011 and May 2014,  two non-overlapping samples of upgoing track-like events, dominated by muons from charged current interactions of atmospheric $\nu_{\mu}$ and $\bar{\nu}_{\mu}$, were isolated \citep{Aartsen:2016zhm}. During austral summers, the Sun being above the horizon, is a source of downgoing neutrinos and the signal is overwhelmed by a background of muons originating in cosmic ray interactions in the upper atmosphere.

The first sample, consisting of events that traverse the principal IceCube array, is sensitive to neutrinos in the 100~GeV -- 1~TeV range in energy, while the second sample is dominated by events starting in and around the DeepCore infill array, and is sensitive down to neutrinos of $\sim$10~GeV in energy. 

An unbinned maximum likelihood ratio analysis of the directions and energies of the events that make up the two samples was unable to identify a statistically significant excess of neutrinos from the direction of the Sun. This enabled 90\% C.L. upper limits on the DM annihilation induced neutrino flux to be computed according to the prescription of ~\cite{Feldman:1997qc} as presented in \cite{Aartsen:2016zhm}.

This can be interpreted as both a constraint on the annihilation rate of DM particles in the Sun, as well as on the scattering cross-section of DM with nucleons, although this has been usually done under the SHM assumption. In particle physics models where the DM couples to the spin of the nucleus and annihilates preferentially into SM particles that decay to produce a large number of high energy neutrinos (such as $\tau^+\tau^-$), the resultant constraints are the most stringent for DM mass above $\sim 80$~GeV~\citep{Tanabashi:2018oca}. 

\bigskip
\bigskip

\subsection{PICO}

The PICO collaboration searches for WIMPs using superheated bubble chambers operated at temperature and pressure conditions which lead to being virtually insensitive to gamma and beta radiation~\citep{Amole:2019scf}. Events in PICO consist of the transition from liquid to gas phase, signalled by the nucleation of a bubble in the target material. This phase change is imaged by the cameras surrounding the active area, which trigger upon detecting the formation of a pocket of gas. Additional background suppression is achieved through the measurement of the acoustic signal generated by the event, allowing alpha particles to be discriminated from nuclear recoils. Details of the apparatus are available in \cite{Amole:2015pla}.  The data used in this study were obtained from the PICO-60 detector, consisting of a 52.2$\pm$0.5~kg C$_\mathrm{3}$ F$_\mathrm{8}$ target, operated roughly two kilometres underground at SNOLAB in Sudbury, Ontario, Canada.  The results used here come from an efficiency-corrected exposure of 1167~kg-days taken between November 2016 and January 2017~\citep{Amole:2017dex}.

The response of the detector to WIMPs is dependent on the thermodynamic conditions, and is calibrated using \textit{in situ} nuclear and electronic recoil sources.  Additionally, the Tandem Van de Graaff facility at the University of Montreal was used to determine the detector response, using well-defined resonances of the $^{51}$V(\textit{p},\textit{n})$^{51}$Cr reaction to produce mono energetic neutrons at 61 and 97~keV.  The combination of these measurements is simulated using differential cross-sections for elastic scattering on fluorine to produce the detector response.

\section{DM velocity distributions and impact on constraints: The method}
\label{sec:method}

Following the method of \cite{Ferrer:2015bta}, the velocity distribution of the DM (WIMP) population in the Solar system, $f(\vec{v})$ can be expressed as the superposition of streams with fixed velocity $\vec{v}_0$ with respect to the Solar frame.

\begin{equation}
f(\vec{v}) = \int_{|\vec{v}_0| \leq v_{\mathrm{max}}} d^3 v_0 \delta^{(3)}(\vec{v} - \vec{v}_0)f(\vec{v}_0)
\label{eq:streamdeco}
\end{equation}

\noindent where $v_{\mathrm{max}}$ is the maximum velocity at which WIMPs can be found, typically the escape velocity of the Galaxy. For every stream with velocity $\vec{v}_0$ with respect to the Sun, upper limits can be derived from the null results of IceCube by requiring that the capture rate for the stream $C_{\vec{v}_0}$ be less than or equal to $ C_{\mathrm{max}}$, the upper limit on the  capture rate from the results of the experiment. For a direct detection experiment, which sees the same stream with velocity $\vec{v}_0 - \vec{v}_{\mathrm{E}}(t)$ with respect to the Earth, similar constraints can be derived for each stream velocity by requiring that the event rate for the stream $R_{\vec{v}_0}$ be less than or equal to $R_{\mathrm{max}}$, the upper limit on the event rate from the results of the experiment. $C_{\vec{v}_0}$ and $R_{\vec{v}_0}$ are computed by evaluating the integrals of equations 2 and 3 of~\cite{Ferrer:2015bta}. Since the PICO exposure period was too short for the Earth's velocity $\vec{v}_{\mathrm{E}}(t)$ to average out to zero, velocities are conservatively shifted by 30.29~km~s$^{-1}$ (the velocity of the Earth around the Sun at perihelion~\citep{er:fa}) when computing $R_{\vec{v}_0}$. For the capture rates in the Sun, the integrals were evaluated using the density profile and nuclear abundances in the Sun for protons and nitrogen nuclei (the second most abundant species with nuclear spin) in the standard Solar model \citep{Bahcall:1987jc} as implemented in \textit{sunpy}~\citep{Sun:Py}.  Nuclear form factors as implemented in \textit{dmdd} \citep{Gluscevic:2015sqa} for spin-dependent scattering, corresponding to the $\Sigma'_{1M}$ (Axial transverse electric response) and $\Sigma''_{1M}$ (Axial longitudinal response), table 1 of ~\cite{Fitzpatrick:2012ix} were employed for the event rate calculations in PICO.

\begin{figure*}
\includegraphics[width=\columnwidth]{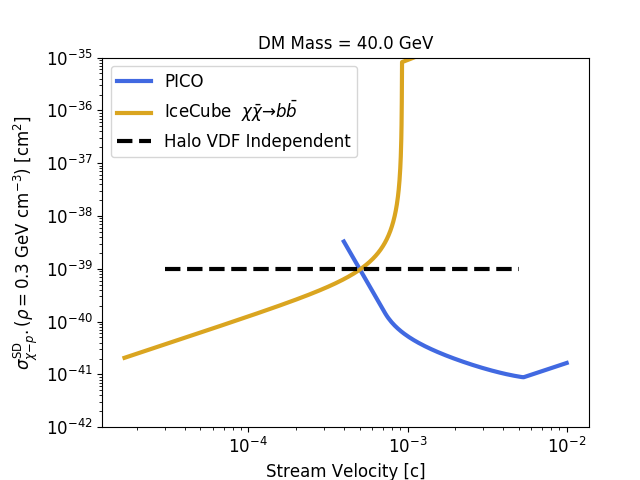} 
\includegraphics[width=\columnwidth]{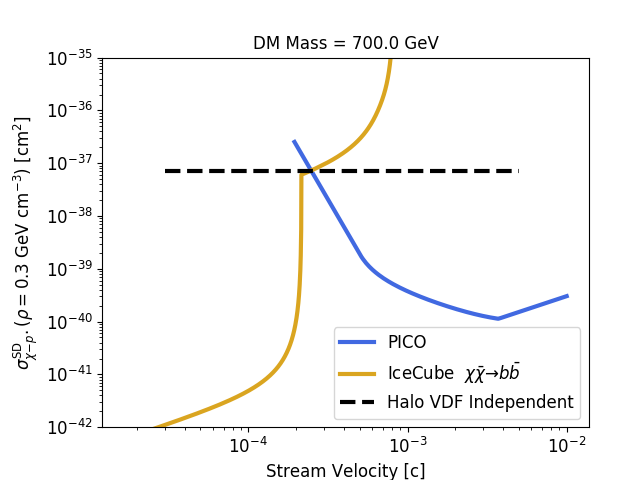} 
\caption{Constraints at $\gtrsim 90\%$ C.L. on the spin-dependent DM-proton scattering cross-section from both IceCube and PICO for different values of $|v_0|$, for 40 (700)~GeV WIMPs annihilating to $b\bar{b}$ are shown on the left (right). For 40~GeV WIMPs, as the efficiency of PICO falls off below stream velocities of $c$ (the speed of light) $\times 10^{-3}$, Solar capture by scattering off hydrogen nuclei provides a complementary bound, while for 700~GeV WIMPs, a bound is provided only by the much less abundant nitrogen nuclei in the Sun.} 
\label{fig:method2specdemo}
\end{figure*}

Figure~\ref{fig:method2specdemo} demonstrates the evolution of the constraints on the spin-dependent DM-proton scattering cross-section from both IceCube and PICO as $|v_0|$ is varied. The individual constraints on the cross section are computed from the constraints on the capture rate in the Sun already derived in \cite{Aartsen:2016zhm} as well as the constraint on the event rate within PICO presented in \cite{Amole:2017dex}. For a WIMP of mass $M$ scattering off a nucleus of mass $m$, the maximum stream velocity at which capture is allowed is given by \citep{Ferrer:2015bta}:

\begin{equation}
v_{\mathrm{max}} = 2 v_{\mathrm{esc}} \sqrt{\frac{Mm}{|M-m|}} 
\label{eq:vmax}
\end{equation}

\noindent where $v_{\mathrm{esc}}$ is the escape velocity. Consequently, above certain threshold values of the stream velocity, capture by scattering off protons is kinematically impossible and only nitrogen nuclei contribute to the capture rate.

Subsequently, the largest value of the scattering cross-section allowed by both IceCube and PICO, $\sigma^{\mathrm{HI}}$, can be determined at the velocity of least constraint, $v_{\mathrm{LC}}$, where $\sigma^{\mathrm{PICO}}_{\mathrm{max}}(v_{\mathrm{LC}}) = \sigma^{\mathrm{IceCube}}_{\mathrm{max}}(v_{\mathrm{LC}})$. This procedure is illustrated in Figure~\ref{fig:method2specdemo} for two specific models, 40~GeV and 700~GeV WIMPs annihilating to $b\bar{b}$.

\bigskip

\section{Results and Conclusions}
\label{sec:results}

\begin{figure*}
\centering
\includegraphics[width=1.8\columnwidth]{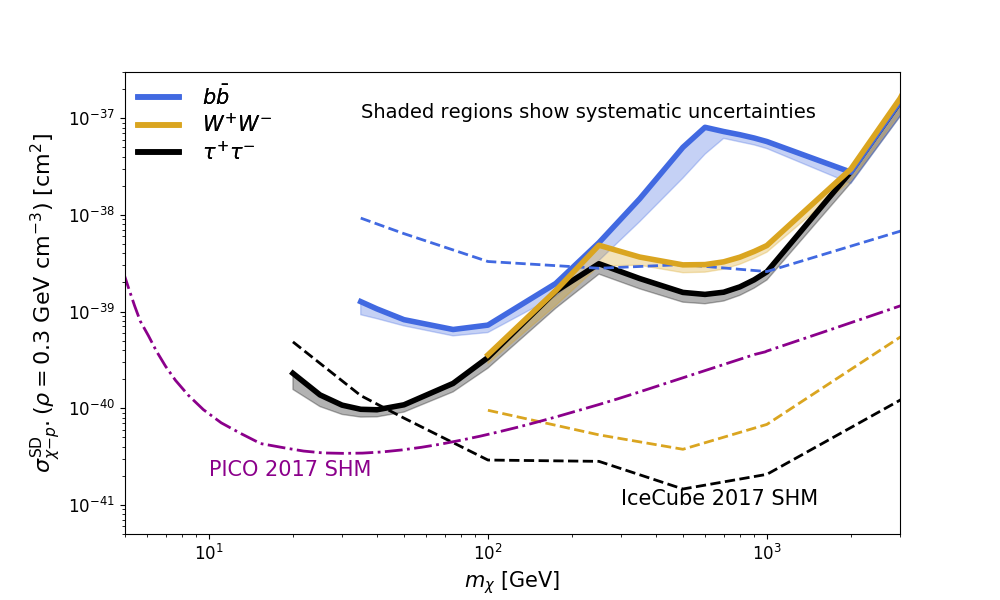} 
\caption{DM velocity distribution independent constraints on the SD DM-nucleon interaction cross-section $\gtrsim 90\%$ C.L. Systematic uncertainties are presented as shaded regions. The traditional SHM upper limits at $90\%$ C.L. from IceCube and PICO are shown as dashed and dash dotted lines. The kinks in the constraints at $\sim 250$~GeV (for \chW and \chtau) and $\sim 700$~GeV (for \chb) are explained in Section \ref{sec:results}.} 
\label{fig:moneyplot}
\end{figure*}

The resultant DM velocity independent constraints are illustrated in Figure~\ref{fig:moneyplot} and presented in Table~\ref{tab:WIMPresults}. For the ``hard" channels ( \chW  and \chtau), which produce a relatively large number of neutrinos at energies just below the DM mass, the DM-velocity-independent constraints are in general worse only by a factor of 2 to 4 compared to the PICO SHM constraints. However, at a DM Mass of $\sim$250~GeV ($\sim$700~GeV for \chb), the constraints are significantly worse because the DM particle velocities just below the PICO threshold are still too high to be captured by scattering off protons in the Sun (see Figure~\ref{fig:method2specdemo}). At immediately higher masses, the constraints improve because the IceCube sensitivity improves with the DM mass in this range. The constraints are in agreement with the findings by \cite{Ibarra:2017mzt}. The IceCube constraints were recomputed with Monte-Carlo data sets under varying assumptions of all systematic uncertainties as described in \cite{Aartsen:2016zhm}. The dominant uncertainties were found to originate in the photodetection efficiency of the photomultiplier tubes that make up the DOMs, as well as the optical properties of the ice. Since these constraints correspond to the same annihilation rates of DM particles in the Sun reported in \cite{Aartsen:2016zhm}, capture-annihilation equilibrium continues to be a valid assumption. The dominant uncertainties in the detector acceptance of PICO originate in the uncertainties of the neutron beam used in the calibration process. These are propagated to the final level and shown as shaded regions. Conservatively, the pessimistic efficiencies of PICO have been used to derive the constraints. While these constraints are robust with respect to any uncertainties in the velocity distribution of DM particles, they are still susceptible to uncertainties and/or fluctuations in the local density of DM, and are presented for the benchmark local density of $\rho_{\mathrm{DM}} = 0.3$~GeV~cm$^{-3}$, and scale inversely with this quantity.

\begin{table*}[]
\centering
\caption{Constraints on the SD DM-nucleon cross-section. SHM constraints from PICO and IceCube, as well as the DM velocity distribution independent constraint are presented at $\gtrsim 90\%$ C.L. The velocity of DM particles at which the cross-section is least constrained ($V_{\mathrm{LC}}$) is also presented for each point. The constraints are conservative with respect to systematic uncertainties.}
\label{tab:WIMPresults}
\begin{tabular}{|l|c|c|c|c|c|c|}
\hline

$m_{\chi}$& annih. & $v_{\mathrm{LC}}$ & PICO & IceCube & Combined & Syst unc. \\
(GeV)& channel & (km s$^{-1})$ &  $\sigma_{\mathrm{SD}}^{\mathrm{SHM}}$ (pb) & $\sigma_{\mathrm{SD}}^{\mathrm{SHM}}$ (pb) & $\sigma_{\mathrm{SD}}^{\mathrm{HI}}$(pb)&(\%) \\
\hline     
20  &  \chtau  &  229.7  &  3.78$\times$10$^{-5}$  &  4.85$\times$10$^{-4}$  &  2.29$\times$10$^{-4}$  &  23.4 \\
\hline
35  &  \chb  &  131.5  &  3.43$\times$10$^{-5}$  &  9.25$\times$10$^{-4}$  &  1.26$\times$10$^{-3}$ &  18.3 \\
35  &  \chtau  &  236.8  &    &  1.35$\times$10$^{-4}$  &  9.74$\times$10$^{-5}$  &  10.2 \\

\hline                                                                                           
50  &  \chb  &  137.3  &  3.72$\times$10$^{-5}$  &  6.39$\times$10$^{-3}$  &  8.24$\times$10$^{-4}$  &  8.0 \\
50  &  \chtau  &  222.5  &   &  7.90$\times$10$^{-5}$  &  1.08$\times$10$^{-4}$  &  9.5 \\
\hline
100  &  \chb  &  141.5  &  &  3.29$\times$10$^{-4}$  &  7.23$\times$10$^{-4}$  &  9.7 \\
100  &  \chW  &  167.8  &  5.36$\times$10$^{-5}$  &  9.52$\times$10$^{-5}$  &  3.56$\times$10$^{-4}$  &  11.4 \\
100  &  \chtau  &  170.3  &  &  2.91$\times$10$^{-5}$  &  3.34$\times$10$^{-4}$  &  14.4 \\
\hline
250  &  \chb  &  106.2  &   &  2.80$\times$10$^{-3}$  &  5.15$\times$10$^{-3}$  &  26.7 \\
250  &  \chW  &  108.3  & 1.09$\times$10$^{-4}$   &  5.30$\times$10$^{-5}$  &  4.85$\times$10$^{-3}$  &  31.8 \\
250  &  \chtau  &  108.4  &    &  2.82$\times$10$^{-5}$  &  3.12$\times$10$^{-3}$  &  14.0 \\
\hline                                  
500  &  \chb  &  76.4  &    &  3.06$\times$10$^{-3}$  &  4.99$\times$10$^{-2}$  &  54.1 \\
500  &  \chW  &  122.7  &  2.06$\times$10$^{-4}$  &  3.76$\times$10$^{-5}$  &  3.04$\times$10$^{-3}$  &  10.2 \\
500  &  \chtau  &  142.5  &    &  1.46$\times$10$^{-5}$  &  1.58$\times$10$^{-3}$  &  13.1 \\
\hline    
1000  &  \chb  &  72.07  &    &  2.59$\times$10$^{-3}$  &  5.72$\times$10$^{-2}$  &  9.1 \\
1000  &  \chW  &  126.0  & 3.90$\times$10$^{-4}$ &  6.80$\times$10$^{-5}$  &  4.81$\times$10$^{-3}$  &  8.6 \\
1000  &  \chtau  &  145.3  &  &  2.07$\times$10$^{-5}$  &  2.57$\times$10$^{-3}$  &  10.8 \\
\hline
3000  &  \chb  &  100.3  &  &  6.76$\times$10$^{-3}$  &  1.61$\times$10$^{-1}$  &  19.8 \\
3000  &  \chW  &  76.09  &  1.14$\times$10$^{-3}$   &  5.42$\times$10$^{-4}$  &  1.59$\times$10$^{-1}$  &  21.4 \\
3000  &  \chtau  &  49.52  &   &  1.21$\times$10$^{-4}$  &  1.48$\times$10$^{-1}$  &  22.4 \\
\hline
5000  &  \chb  &  89.23  &   &  1.58$\times$10$^{-2}$  &  3.11  & 25.4   \\
5000  &  \chW  &  46.41  & 1.89$\times$10$^{-3}$  &  1.37$\times$10$^{-3}$  &  3.16  & 16.5 \\
5000  &  \chtau  &  46.41  &    &  3.28$\times$10$^{-4}$  &  2.66  & 19.1 \\

\bottomrule
\end{tabular}
\end{table*}

\begin{acknowledgments}
\underline{\textbf{Acknowledgements:}}

We would like to acknowledge valuable discussions with John Ellis and Alejandro Ibarra about the analysis.

\textbf{IceCube Collaboration:}

USA {\textendash} U.S. National Science Foundation-Office of Polar Programs,
U.S. National Science Foundation-Physics Division,
Wisconsin Alumni Research Foundation,
Center for High Throughput Computing (CHTC) at the University of Wisconsin-Madison,
Open Science Grid (OSG),
Extreme Science and Engineering Discovery Environment (XSEDE),
U.S. Department of Energy-National Energy Research Scientific Computing Center,
Particle astrophysics research computing center at the University of Maryland,
Institute for Cyber-Enabled Research at Michigan State University,
and Astroparticle physics computational facility at Marquette University;
Belgium {\textendash} Funds for Scientific Research (FRS-FNRS and FWO),
FWO Odysseus and Big Science programmes,
and Belgian Federal Science Policy Office (Belspo);
Germany {\textendash} Bundesministerium f{\"u}r Bildung und Forschung (BMBF),
Deutsche Forschungsgemeinschaft (DFG),
Helmholtz Alliance for Astroparticle Physics (HAP),
Initiative and Networking Fund of the Helmholtz Association,
Deutsches Elektronen Synchrotron (DESY),
and High Performance Computing cluster of the RWTH Aachen;
Sweden {\textendash} Swedish Research Council,
Swedish Polar Research Secretariat,
Swedish National Infrastructure for Computing (SNIC),
and Knut and Alice Wallenberg Foundation;
Australia {\textendash} Australian Research Council;
Canada {\textendash} Natural Sciences and Engineering Research Council of Canada,
Calcul Qu{\'e}bec, Compute Ontario, Canada Foundation for Innovation, WestGrid, and Compute Canada;
Denmark {\textendash} Villum Fonden, Danish National Research Foundation (DNRF), Carlsberg Foundation;
New Zealand {\textendash} Marsden Fund;
Japan {\textendash} Japan Society for Promotion of Science (JSPS)
and Institute for Global Prominent Research (IGPR) of Chiba University;
Korea {\textendash} National Research Foundation of Korea (NRF);
Switzerland {\textendash} Swiss National Science Foundation (SNSF).
United Kingdom {\textendash} Department of Physics, University of Oxford.

\textbf{PICO Collaboration:}

The PICO collaboration wishes to thank SNOLAB and its staff for support through underground space, logistical and technical services. SNOLAB operations are supported by the Canada Foundation for Innovation and the Province of Ontario Ministry of Research and Innovation, with underground access provided by Vale at the Creighton mine site. We wish to acknowledge the support of the Natural Sciences and Engineering Research Council of Canada (NSERC) and the Canada Foundation for Innovation (CFI) for funding. We acknowledge the support from the National Science Foundation (NSF) (Grant 0919526, 1506337, 1242637 and 1205987). We acknowledge that this work is supported by the U.S. Department of Energy (DOE) Office of Science, Office of High Energy Physics (under award DE-SC-0012161), by the DOE Office of Science Graduate Student Research (SCGSR) award, by DGAPA-UNAM (PAPIIT No. IA100118) and Consejo Nacional de Ciencia y Tecnoloǵıa (CONACyT, M\'exico, Grant No. 252167  and A1-S-8960), by the Department of Atomic Energy (DAE), Government of India, under the Centre for AstroParticle Physics II project (CAPP-II) at the Saha Institute of Nuclear Physics (SINP), European Regional Development Fund- Project “Engineering applications of microworld physics” (No. CZ.02.1.01/0.0/0.0/16 019/0000766), and the Spanish Ministerio de Ciencia, Innovaci\'on y Universidades (Red Consolider MultiDark, FPA2017--90566--REDC). This work is partially supported by the Kavli Institute for Cosmological Physics at the University of Chicago through NSF grant 1125897 and 1806722, and an endowment from the Kavli Foundation and its founder Fred Kavli. We also wish to acknowledge the support from Fermi National Accelerator Laboratory under Contract No.DE-AC02-07CH11359, and from Pacific Northwest National Laboratory, which is operated by Battelle for the U.S. Department of Energy under Contract No. DE- AC05-76RL01830. We also thank Compute Canada (www.computecanada.ca) and the Centre for Advanced Computing, ACENET, Calcul Quebe\'c, Compute Ontario and WestGrid for computational support.
\end{acknowledgments}


\end{document}